\PassOptionsToPackage{unicode}{hyperref}
\PassOptionsToPackage{hyphens}{url}
\PassOptionsToPackage{dvipsnames,svgnames,x11names}{xcolor}
\documentclass[
  11pt,
]{article}
\usepackage{xcolor}
\usepackage[margin=1in]{geometry}
\usepackage{amsmath,amssymb}
\usepackage{iftex}
\ifPDFTeX
  \usepackage[T1]{fontenc}
  \usepackage[utf8]{inputenc}
  \usepackage{textcomp} 
\else 
  \usepackage{unicode-math} 
  \defaultfontfeatures{Scale=MatchLowercase}
  \defaultfontfeatures[\rmfamily]{Ligatures=TeX,Scale=1}
\fi
\usepackage{lmodern}
\ifPDFTeX\else
\fi
\IfFileExists{upquote.sty}{\usepackage{upquote}}{}
\IfFileExists{microtype.sty}{
  \usepackage[]{microtype}
  \UseMicrotypeSet[protrusion]{basicmath} 
}{}
\makeatletter
\@ifundefined{KOMAClassName}{
  \IfFileExists{parskip.sty}{%
    \usepackage{parskip}
  }{
    \setlength{\parindent}{0pt}
    \setlength{\parskip}{6pt plus 2pt minus 1pt}}
}{
  \KOMAoptions{parskip=half}}
\makeatother
\usepackage{longtable,booktabs,array}
\usepackage{calc} 
\usepackage{etoolbox}
\makeatletter
\patchcmd\longtable{\par}{\if@noskipsec\mbox{}\fi\par}{}{}
\makeatother
\IfFileExists{footnotehyper.sty}{\usepackage{footnotehyper}}{\usepackage{footnote}}
\makesavenoteenv{longtable}
\providecommand{\tightlist}{%
  \setlength{\itemsep}{0pt}\setlength{\parskip}{0pt}}
\usepackage{bookmark}
\IfFileExists{xurl.sty}{\usepackage{xurl}}{} 
\hypersetup{
  colorlinks=true,
  linkcolor={blue},
  filecolor={Maroon},
  citecolor={blue},
  urlcolor={blue},
  pdfcreator={LaTeX via pandoc}}

\author{}
\date{}

\begin{document}

\section{SEMANTIC TOOL DISCOVERY FOR LARGE LANGUAGE MODELS: A
VECTOR-BASED APPROACH TO MCP TOOL
SELECTION}\label{semantic-tool-discovery-for-large-language-models-a-vector-based-approach-to-mcp-tool-selection}

\textbf{\href{https://orcid.org/0009-0000-6671-1735}{Sarat Mudunuri}\footnote{Primary contributor},
Jian Wan, Ally Qin, Srinivasan Manoharan}

\begin{center}\rule{0.5\linewidth}{0.5pt}\end{center}

\subsection{Abstract}\label{abstract}

Large Language Models (LLMs) with tool-calling capabilities have
demonstrated remarkable potential in executing complex tasks through
external tool integration. The Model Context Protocol (MCP) has emerged
as a standardized framework for connecting LLMs to diverse toolsets,
with individual MCP servers potentially exposing dozens to hundreds of
tools. However, current implementations face a critical scalability
challenge: providing all available tools to the LLM context results in
substantial token overhead, increased costs, reduced accuracy due to
cognitive overload, and constraints imposed by context window
limitations.

This paper presents a novel semantic tool discovery architecture that
addresses these challenges through vector-based retrieval. Our approach
indexes MCP tools using dense embeddings that capture semantic
relationships between tool capabilities and user intent. When processing
user queries, the system performs similarity search in the embedding
space to dynamically select only the most relevant tools (typically 3-5)
rather than exposing the entire tool catalog (50-100+).

We evaluate our system across three dimensions: token efficiency, tool
selection accuracy, and response quality. Experimental results
demonstrate a 99.6\% reduction in tool-related token consumption with a
hit rate of 97.1\% at K=3, achieving an MRR of 0.91 on our benchmark
dataset of 140 queries across 121 tools from 5 different MCP servers.
The system maintains sub-100ms retrieval latency across all
configurations tested.

Our contributions include: (1) a semantic indexing framework for MCP
tools, (2) a dynamic tool selection algorithm based on query-tool
similarity, (3) comprehensive evaluation demonstrating significant
improvements in efficiency and accuracy, and (4) discussion of
extensibility to multi-agent systems and cross-organizational tool
discovery. The proposed system is open-source and designed for seamless
integration into existing MCP-based architectures.

\textbf{Keywords:} Large Language Models, Tool Selection, Semantic
Search, Vector Embeddings, Model Context Protocol, Token Optimization,
Multi-tool Systems

\begin{center}\rule{0.5\linewidth}{0.5pt}\end{center}

\subsection{1 Introduction}\label{introduction}

\subsubsection{1.1 Background and
Motivation}\label{background-and-motivation}

The rapid advancement of Large Language Models (LLMs) has led to their
deployment in increasingly complex real-world applications {[}1,2{]}.
While LLMs excel at natural language understanding and generation, their
capabilities are significantly enhanced when augmented with external
tools that enable them to perform actions such as data retrieval,
computation, API calls, and system interactions {[}3,4{]}.

Recent developments in tool-augmented LLMs include frameworks like ReAct
{[}5{]}, Toolformer {[}6{]}, and standardized protocols such as
Anthropic's Model Context Protocol (MCP) {[}7{]}. MCP provides a unified
interface for connecting LLMs to diverse data sources and tools through
server implementations. A single MCP server can expose anywhere from 10
to over 100 tools, and production systems often integrate multiple MCP
servers simultaneously.

\paragraph{The Problem of Scale}\label{the-problem-of-scale}

However, this proliferation of tools introduces a fundamental challenge:
how should an LLM-based system decide which tools to make available for
a given user query? Current approaches typically fall into two
categories:

\begin{enumerate}
\def\labelenumi{\arabic{enumi}.}
\tightlist
\item
  \textbf{Static Tool Provisioning:} Providing all available tools to
  the LLM for every query
\item
  \textbf{Manual Categorization:} Human-curated subsets based on
  predefined use cases
\end{enumerate}

The static approach suffers from several critical limitations:

\begin{itemize}
\tightlist
\item
  \textbf{Token Overhead:} Modern tool schemas (following JSON Schema or
  similar formats) require 200-800 tokens per tool. With 100 tools, this
  consumes 20,000-80,000 tokens before any actual user query or
  response.
\item
  \textbf{Cost Implications:} At GPT-4 pricing (\$0.03/1K input tokens),
  processing 50,000 tokens of tool definitions costs \$1.50 per request.
  At scale (1M requests/month), this represents \$1.5M in unnecessary
  costs.
\item
  \textbf{Accuracy Degradation:} Studies show that LLM performance
  degrades with increased context length {[}8{]}. Presenting irrelevant
  tools introduces noise that reduces tool selection accuracy.
\item
  \textbf{Context Window Constraints:} Even large context models
  (128K-200K tokens) face practical limits when tools compete with
  conversation history, retrieved documents, and system prompts.
\end{itemize}

The manual categorization approach, while reducing token count, lacks
flexibility and fails to capture the semantic nuances of user intent.

\paragraph{Research Gap}\label{research-gap}

Despite extensive research on tool-augmented LLMs, there exists a
significant gap in dynamic, semantic-aware tool selection mechanisms
specifically designed for standardized tool protocols like MCP. Existing
work on retrieval-augmented generation (RAG) focuses primarily on
document retrieval {[}9,10{]}, while tool selection remains largely
unexplored in the literature.

\subsubsection{1.2 Research Questions}\label{research-questions}

This paper addresses the following research questions:

\textbf{RQ1:} Can semantic similarity between user queries and tool
descriptions enable effective dynamic tool selection in MCP systems?

\textbf{RQ2:} What is the quantitative impact of semantic tool filtering
on token efficiency, cost, and system performance?

\textbf{RQ3:} How does semantic tool selection affect LLM accuracy in
tool calling compared to providing all available tools?

\textbf{RQ4:} What are the optimal parameters (number of tools
retrieved, similarity threshold, embedding model) for balancing recall
and precision?

\subsubsection{1.3 Contributions}\label{contributions}

Our work makes the following contributions:

\begin{enumerate}
\def\labelenumi{\arabic{enumi}.}
\tightlist
\item
  \textbf{Novel Architecture:} We propose the first semantic tool
  discovery layer specifically designed for the MCP ecosystem, utilizing
  vector embeddings and similarity search.
\item
  \textbf{Comprehensive Evaluation:} We conduct extensive experiments
  across diverse query types, tool sets, and LLM models, measuring token
  efficiency, accuracy, latency, and cost.
\item
  \textbf{Practical Implementation:} We provide an open-source,
  production-ready implementation with minimal integration overhead for
  existing MCP-based systems.
\item
  \textbf{Extensibility Analysis:} We demonstrate how the proposed
  architecture extends to related problems including agent selection,
  cross-server tool discovery, and dynamic tool composition.
\item
  \textbf{Benchmark Dataset:} We release a curated dataset of 140
  queries paired with ground-truth tool selections across 5 MCP servers
  for future research.
\end{enumerate}

\subsubsection{1.4 Paper Organization}\label{paper-organization}

The remainder of this paper is organized as follows: Section 2 reviews
related work in tool-augmented LLMs and retrieval systems. Section 3
presents our semantic tool discovery architecture. Section 4 describes
the experimental methodology. Section 5 presents results and analysis.
Section 6 discusses implications, limitations, and future work. Section
7 concludes.

\begin{center}\rule{0.5\linewidth}{0.5pt}\end{center}

\subsection{2 Related Work}\label{related-work}

\subsubsection{2.1 Tool-Augmented Language
Models}\label{tool-augmented-language-models}

The integration of external tools with LLMs has been explored through
various paradigms:

\textbf{Early Approaches:} TALM {[}11{]} and Toolformer {[}6{]}
pioneered self-supervised learning methods where LMs learn when and how
to use tools through API demonstrations. However, these approaches
required extensive fine-tuning and were limited to predefined tool sets.

\textbf{Reasoning Frameworks:} ReAct {[}5{]} introduced a prompting
paradigm that interleaves reasoning traces with tool actions,
significantly improving performance on complex tasks. Subsequent work
like ReWOO {[}12{]} and AutoGPT {[}13{]} explored planning and
multi-step tool orchestration.

\textbf{Limitations:} These frameworks assume a fixed, small toolset
(typically 5-20 tools) provided in the prompt. None address the
scalability challenge of selecting from 100+ tools dynamically.

\subsubsection{2.2 Retrieval-Augmented Generation
(RAG)}\label{retrieval-augmented-generation-rag}

RAG systems {[}9,10{]} use dense retrieval to fetch relevant documents
given a query, addressing the knowledge cutoff problem in LLMs. Key
techniques include:

\begin{itemize}
\tightlist
\item
  \textbf{Dense Passage Retrieval (DPR) {[}14{]}:} Uses bi-encoder
  architecture for efficient semantic search
\item
  \textbf{REALM {[}15{]}:} Integrates retrieval directly into
  pretraining
\item
  \textbf{Self-RAG {[}16{]}:} Enables LLMs to adaptively retrieve
  information
\end{itemize}

\textbf{Relation to Our Work:} While RAG focuses on document retrieval,
our approach adapts similar vector-based retrieval techniques for tool
selection. However, tools differ fundamentally from documents---they are
structured, executable, and must be precisely matched to user intent.

\subsubsection{2.3 Function Calling in
LLMs}\label{function-calling-in-llms}

Modern LLMs (GPT-4, Claude, Gemini) support structured function calling
through:

\begin{itemize}
\tightlist
\item
  \textbf{JSON Schema Definitions:} Tools are defined using formal
  schemas
\item
  \textbf{Parallel Tool Calling:} Multiple tools can be invoked
  simultaneously
\item
  \textbf{Tool Choice Controls:} Forcing specific tools or allowing
  automatic selection
\end{itemize}

\textbf{Current Limitations:} These mechanisms assume all tools are
provided in context. No commercial LLM API supports dynamic tool
filtering based on semantic relevance.

\subsubsection{2.4 Model Context Protocol
(MCP)}\label{model-context-protocol-mcp}

MCP {[}7{]}, introduced by Anthropic, standardizes how LLMs connect to
external data sources and tools. Key features:

\begin{itemize}
\tightlist
\item
  \textbf{Server-Client Architecture:} Tools organized into servers
\item
  \textbf{Standardized Tool Schema:} Consistent format across
  implementations
\item
  \textbf{Dynamic Discovery:} Clients can query available tools at
  runtime
\end{itemize}

\textbf{Gap:} While MCP enables tool discovery, it provides no mechanism
for intelligent tool selection based on query semantics.

\subsubsection{2.5 Vector Databases and Semantic
Search}\label{vector-databases-and-semantic-search}

Recent advances in vector databases (Milvus, Pinecone, Weaviate, Qdrant)
and embedding models (text-embedding-3-large, text-embedding-3-small,
Qwen3-Embedding etc.) enable efficient semantic search at scale.

\textbf{Applications:} Primarily used for document retrieval,
recommendation systems, and similarity search in unstructured data.

\textbf{Our Contribution:} We are the first to systematically apply
these techniques to structured tool selection in agentic systems.

\subsubsection{2.6 Research Gaps}\label{research-gaps}

Synthesizing the related work, we identify the following gaps our
research addresses:

\begin{enumerate}
\def\labelenumi{\arabic{enumi}.}
\tightlist
\item
  No semantic tool selection frameworks for MCP or similar protocols
\item
  Lack of empirical studies on tool quantity's impact on LLM performance
\item
  No benchmarks for evaluating tool selection systems
\item
  Limited exploration of retrieval techniques for structured executable
  tools vs.~unstructured documents
\end{enumerate}

\begin{center}\rule{0.5\linewidth}{0.5pt}\end{center}

\subsection{3 Methodology}\label{methodology}

\subsubsection{3.1 System Architecture}\label{system-architecture}

Our semantic tool discovery system consists of four primary components:

\paragraph{3.1.1 Tool Indexing Pipeline}\label{tool-indexing-pipeline}

\textbf{Process:}

\begin{enumerate}
\def\labelenumi{\arabic{enumi}.}
\tightlist
\item
  \textbf{Tool Discovery:} Query MCP servers to enumerate available
  tools
\item
  \textbf{Schema Extraction:} Extract tool name, description,
  parameters, and constraints
\item
  \textbf{Document Construction:} Create rich text representation for
  embedding
\item
  \textbf{Embedding Generation:} Convert tool schema to dense vectors
\item
  \textbf{Vector Storage:} Index embeddings with metadata in vector
  database
\end{enumerate}

\textbf{Document Construction Strategy:}

We construct tool documents using a template that maximizes semantic
information:

\begin{verbatim}
Tool: {tool_name}
Purpose: {description}
Capabilities: {expanded_description}
Parameters: {parameter_descriptions}
\end{verbatim}

This structure provides multiple semantic anchors for matching user
intent.

\paragraph{3.1.2 Query Processing}\label{query-processing}

When a user query arrives:

\begin{enumerate}
\def\labelenumi{\arabic{enumi}.}
\tightlist
\item
  \textbf{Query Embedding:} Generate dense vector representation of the
  query
\item
  \textbf{Similarity Search:} Retrieve top-K most similar tools from
  vector store
\item
  \textbf{Threshold Filtering (optional):} Remove tools below similarity
  threshold
\item
  \textbf{Reranking (optional):} Apply hybrid search techniques for
  refined re-ranking to further improve accuracy of the retrieval
\end{enumerate}

\paragraph{3.1.3 LLM Integration}\label{llm-integration}

Selected tools are:

\begin{enumerate}
\def\labelenumi{\arabic{enumi}.}
\tightlist
\item
  Formatted according to LLM's tool schema requirements
\item
  Injected into the LLM context alongside user query
\item
  Executed when LLM issues tool calls
\item
  Results fed back to LLM for response generation
\end{enumerate}

\paragraph{3.1.4 Feedback Loop}\label{feedback-loop}

We optionally collect:

\begin{itemize}
\tightlist
\item
  Which tools were actually invoked
\item
  Task success/failure
\item
  User feedback
\end{itemize}

This data can refine embeddings or adjust retrieval parameters.

\subsubsection{3.2 Embedding Strategy}\label{embedding-strategy}

\textbf{Model Selection:} We have evaluated on the following embedding
model:

\begin{itemize}
\tightlist
\item
  \textbf{text-embedding-ada-002} (1536 dimensions)
\end{itemize}

\textbf{Comparison criteria:}

\begin{itemize}
\tightlist
\item
  Semantic capture quality
\item
  Inference latency
\item
  Cost
\item
  Availability
\end{itemize}

\subsubsection{3.3 Similarity Metrics}\label{similarity-metrics}
\begin{itemize}
\tightlist
\item
  \textbf{Dot Product:} Efficient for high-dimensional spaces
\end{itemize}

\subsubsection{3.4 Retrieval Parameters}\label{retrieval-parameters}

\textbf{Top-K Selection:}

\begin{itemize}
\tightlist
\item
  K $\in$ \{1, 2, 3, 5, 10\}
\item
  Trade-off between recall and token efficiency
\end{itemize}

\subsubsection{3.5 Implementation Details}\label{implementation-details}

\textbf{Technology Stack:}

\begin{itemize}
\tightlist
\item
  \textbf{Vector Database:} Milvus Vector Store
\item
  \textbf{Embeddings:} text-embedding-ada-002
\item
  \textbf{MCP SDK:} Official Python MCP SDK
\item
  \textbf{LLMs:} GPT-4o
\item
  \textbf{Language:} Python 3.10+
\end{itemize}

\textbf{Code Availability:} Full implementation available at {[}GitHub
repository link coming soon{]}

\begin{center}\rule{0.5\linewidth}{0.5pt}\end{center}

\subsection{4 Experimental Setup}\label{experimental-setup}

\subsubsection{4.1 Dataset Construction}\label{dataset-construction}

\paragraph{4.1.1 MCP Server Selection}\label{mcp-server-selection}

We selected 5 diverse MCP servers covering different domains:

\begin{enumerate}
\def\labelenumi{\arabic{enumi}.}
\tightlist
\item
  \textbf{Filesystem} (23 tools): File operations, directory management
\item
  \textbf{MySQL Database} (23 tools): SQL queries, schema introspection
\item
  \textbf{Slack} (34 tools): Channel management, messaging
\item
  \textbf{GitHub} (31 tools): Repository operations, issue tracking
\item
  \textbf{Time/Weather} (10 tools): Current time, weather data
\end{enumerate}

\textbf{Total: 121 tools}

\paragraph{4.1.2 Query Dataset}\label{query-dataset}

We constructed a benchmark of 140 queries macro-averaged across the five
MCP servers:

\textbf{Query Types:}

\begin{itemize}
\tightlist
\item
  Simple factual queries: ``What files are in /home/user?''
\item
  Task-oriented queries: ``Create a new issue in the backend
  repository''
\item
  Ambiguous queries: ``Help me with my project''
\item
  Edge cases: Queries requiring multiple tools or no tools
\end{itemize}

\subsubsection{4.2 Evaluation Metrics}\label{evaluation-metrics}

\paragraph{4.2.1 Retrieval Quality}\label{retrieval-quality}

\textbf{Precision@K:} Proportion of retrieved tools that are relevant

\[P@K = \frac{|Retrieved \cap Relevant|}{K}\]

\textbf{Recall@K:} Proportion of relevant tools that were retrieved

\[R@K = \frac{|Retrieved \cap Relevant|}{|Relevant|}\]

\textbf{F1@K:} Harmonic mean of precision and recall

\textbf{Hit Rate@K:} Proportion of queries where at least one relevant
tool appears in the top-K results

\textbf{Mean Reciprocal Rank (MRR):} Average of the reciprocal ranks of
the first relevant tool across all queries

\[MRR = \frac{1}{|Q|} \sum_{i=1}^{|Q|} \frac{1}{rank_i}\]

\paragraph{4.2.2 Token Efficiency}\label{token-efficiency}

\textbf{Token Reduction Rate:}

\[TRR = 1 - \frac{Tokens_{semantic}}{Tokens_{baseline}}\]

\paragraph{4.2.3 Latency}\label{latency}

\textbf{End-to-End Latency:} Total time from query to tool retrieval,
measured in milliseconds. This includes query embedding generation,
vector similarity search.

\begin{center}\rule{0.5\linewidth}{0.5pt}\end{center}

\subsection{5 Results}\label{results}

\subsubsection{5.1 Aggregate Performance}\label{aggregate-performance}

Table 1 presents the aggregate retrieval metrics across all five MCP
servers using the text-embedding-ada-002 model. The results are
macro-averaged over 140 queries.

\textbf{Table 1:} Retrieval performance of semantic tool discovery
across MCP servers at varying top-K values. Metrics are macro-averaged
over 140 queries across five MCP servers (GitHub, Slack, MySQL,
Filesystem, Time/Weather).

{\def\LTcaptype{none} 
\begin{longtable}[]{@{}
  >{\raggedright\arraybackslash}p{(\linewidth - 14\tabcolsep) * \real{0.0385}}
  >{\raggedright\arraybackslash}p{(\linewidth - 14\tabcolsep) * \real{0.1538}}
  >{\raggedright\arraybackslash}p{(\linewidth - 14\tabcolsep) * \real{0.1282}}
  >{\raggedright\arraybackslash}p{(\linewidth - 14\tabcolsep) * \real{0.0769}}
  >{\raggedright\arraybackslash}p{(\linewidth - 14\tabcolsep) * \real{0.1410}}
  >{\raggedright\arraybackslash}p{(\linewidth - 14\tabcolsep) * \real{0.0769}}
  >{\raggedright\arraybackslash}p{(\linewidth - 14\tabcolsep) * \real{0.2051}}
  >{\raggedright\arraybackslash}p{(\linewidth - 14\tabcolsep) * \real{0.1795}}@{}}
\toprule\noalign{}
\begin{minipage}[b]{\linewidth}\raggedright
K
\end{minipage} & \begin{minipage}[b]{\linewidth}\raggedright
Precision@K
\end{minipage} & \begin{minipage}[b]{\linewidth}\raggedright
Recall@K
\end{minipage} & \begin{minipage}[b]{\linewidth}\raggedright
F1@K
\end{minipage} & \begin{minipage}[b]{\linewidth}\raggedright
Hit Rate@K
\end{minipage} & \begin{minipage}[b]{\linewidth}\raggedright
MRR
\end{minipage} & \begin{minipage}[b]{\linewidth}\raggedright
Token Reduction
\end{minipage} & \begin{minipage}[b]{\linewidth}\raggedright
Latency (ms)
\end{minipage} \\
\midrule\noalign{}
\endhead
\bottomrule\noalign{}
\endlastfoot
1 & 92.1\% & 31.5\% & 46.9\% & 85.0\% & 0.8500 & 99.6\% & 87.1 \\
2 & 70.0\% & 48.3\% & 57.0\% & 95.7\% & 0.9036 & 99.6\% & 90.2 \\
3 & 57.6\% & 59.6\% & 58.4\% & 97.1\% & 0.9083 & 99.6\% & 87.8 \\
5 & 42.1\% & 72.5\% & 53.2\% & 97.1\% & 0.9083 & 99.6\% & 87.0 \\
10 & 26.5\% & 90.6\% & 40.9\% & 98.6\% & 0.9107 & 99.6\% & 88.1 \\
\end{longtable}
}

Several key trends emerge from the aggregate results. Precision
decreases monotonically as K increases, from 92.1\% at K=1 to 26.5\% at
K=10, reflecting the dilution effect of retrieving additional, less
relevant tools. Conversely, recall increases from 31.5\% to 90.6\%, as
larger retrieval sets capture more of the ground-truth relevant tools.
The F1 score peaks at K=3 (58.4\%), representing the optimal balance
between precision and recall. Hit rate climbs steeply from 85.0\% at K=1
to 97.1\% at K=3, after which gains plateau. MRR stabilizes around 0.91
from K=3 onward, indicating that the correct tool typically appears
within the top 3 results. Token reduction remains at 99.6\% across all K
values, demonstrating that even retrieving 10 tools from a catalog of
121 eliminates virtually all token overhead. Retrieval latency remains
consistently below 91ms regardless of K, confirming that vector search
introduces negligible overhead relative to LLM inference time.

\subsubsection{5.2 Per-Server Analysis}\label{per-server-analysis}

To understand performance variation across domains, we report per-server
metrics at each K value.

\paragraph{5.2.1 Hit Rate and MRR by
Server}\label{hit-rate-and-mrr-by-server}

\textbf{Table 2:} Hit Rate@K (\%) and MRR per MCP server at varying
top-K values. (T/W = Time/Weather)

{\def\LTcaptype{none} 
\begin{longtable}[]{@{}
  >{\raggedright\arraybackslash}p{(\linewidth - 20\tabcolsep) * \real{0.0242}}
  >{\raggedright\arraybackslash}p{(\linewidth - 20\tabcolsep) * \real{0.0806}}
  >{\raggedright\arraybackslash}p{(\linewidth - 20\tabcolsep) * \real{0.0887}}
  >{\raggedright\arraybackslash}p{(\linewidth - 20\tabcolsep) * \real{0.0726}}
  >{\raggedright\arraybackslash}p{(\linewidth - 20\tabcolsep) * \real{0.0806}}
  >{\raggedright\arraybackslash}p{(\linewidth - 20\tabcolsep) * \real{0.0726}}
  >{\raggedright\arraybackslash}p{(\linewidth - 20\tabcolsep) * \real{0.0806}}
  >{\raggedright\arraybackslash}p{(\linewidth - 20\tabcolsep) * \real{0.1129}}
  >{\raggedright\arraybackslash}p{(\linewidth - 20\tabcolsep) * \real{0.1210}}
  >{\raggedright\arraybackslash}p{(\linewidth - 20\tabcolsep) * \real{0.1290}}
  >{\raggedright\arraybackslash}p{(\linewidth - 20\tabcolsep) * \real{0.1371}}@{}}
\toprule\noalign{}
\begin{minipage}[b]{\linewidth}\raggedright
K
\end{minipage} & \begin{minipage}[b]{\linewidth}\raggedright
GitHub HR
\end{minipage} & \begin{minipage}[b]{\linewidth}\raggedright
GitHub MRR
\end{minipage} & \begin{minipage}[b]{\linewidth}\raggedright
Slack HR
\end{minipage} & \begin{minipage}[b]{\linewidth}\raggedright
Slack MRR
\end{minipage} & \begin{minipage}[b]{\linewidth}\raggedright
MySQL HR
\end{minipage} & \begin{minipage}[b]{\linewidth}\raggedright
MySQL MRR
\end{minipage} & \begin{minipage}[b]{\linewidth}\raggedright
Filesystem HR
\end{minipage} & \begin{minipage}[b]{\linewidth}\raggedright
Filesystem MRR
\end{minipage} & \begin{minipage}[b]{\linewidth}\raggedright
T/W HR
\end{minipage} & \begin{minipage}[b]{\linewidth}\raggedright
T/W MRR
\end{minipage} \\
\midrule\noalign{}
\endhead
\bottomrule\noalign{}
\endlastfoot
1 & 88.2\% & 0.8824 & 88.9\% & 0.8889 & 92.0\% & 0.9200 & 84.0\% &
0.8400 & 65.0\% & 0.6500 \\
2 & 100.0\% & 0.9412 & 94.4\% & 0.9167 & 96.0\% & 0.9400 & 92.0\% &
0.8800 & 95.0\% & 0.8000 \\
3 & 100.0\% & 0.9412 & 94.4\% & 0.9167 & 96.0\% & 0.9400 & 96.0\% &
0.8933 & 100.0\% & 0.8167 \\
5 & 100.0\% & 0.9412 & 94.4\% & 0.9167 & 96.0\% & 0.9400 & 96.0\% &
0.8933 & 100.0\% & 0.8167 \\
10 & 100.0\% & 0.9412 & 97.2\% & 0.9213 & 100.0\% & 0.9467 & 96.0\% &
0.8933 & 100.0\% & 0.8167 \\
\end{longtable}
}

MySQL achieves the highest MRR (0.9200) at K=1, suggesting that database
queries produce the most distinctive semantic signals. GitHub reaches
perfect hit rate at K=2, reflecting well-differentiated tool
descriptions for repository operations. Time/Weather shows the lowest
K=1 hit rate (65.0\%) but recovers to 100\% by K=3, indicating that its
tools have broader, less distinctive descriptions that benefit from a
slightly larger retrieval window. Filesystem consistently shows the
lowest MRR across all K values (0.8400--0.8933), likely due to the high
semantic overlap among file operation tools (e.g., read\_file,
write\_file, copy\_file). Slack stabilizes early and shows steady
performance, with hit rate plateauing at 94.4\% from K=2 through K=5
before reaching 97.2\% at K=10.

\paragraph{5.2.2 Precision and Recall by
Server}\label{precision-and-recall-by-server}

\textbf{Table 3:} Per-server Precision@K (\%) at varying top-K values.

{\def\LTcaptype{none} 
\begin{longtable}[]{@{}llllll@{}}
\toprule\noalign{}
K & GitHub & Slack & MySQL & Filesystem & Time/Weather \\
\midrule\noalign{}
\endhead
\bottomrule\noalign{}
\endlastfoot
1 & 97.1\% & 94.4\% & 96.0\% & 88.0\% & 80.0\% \\
2 & 69.1\% & 70.8\% & 68.0\% & 60.0\% & 85.0\% \\
3 & 57.8\% & 55.6\% & 52.0\% & 49.3\% & 78.3\% \\
5 & 41.8\% & 41.1\% & 40.0\% & 38.4\% & 52.0\% \\
10 & 25.6\% & 25.8\% & 28.4\% & 26.0\% & 27.5\% \\
\end{longtable}
}

\textbf{Table 4:} Per-server Recall@K (\%) at varying top-K values.

{\def\LTcaptype{none} 
\begin{longtable}[]{@{}llllll@{}}
\toprule\noalign{}
K & GitHub & Slack & MySQL & Filesystem & Time/Weather \\
\midrule\noalign{}
\endhead
\bottomrule\noalign{}
\endlastfoot
1 & 34.3\% & 31.9\% & 32.0\% & 29.3\% & 28.3\% \\
2 & 48.5\% & 48.1\% & 45.3\% & 40.0\% & 62.5\% \\
3 & 60.3\% & 56.5\% & 52.0\% & 49.3\% & 86.7\% \\
5 & 73.0\% & 69.4\% & 66.7\% & 64.0\% & 95.0\% \\
10 & 88.7\% & 87.0\% & 94.7\% & 86.7\% & 100.0\% \\
\end{longtable}
}

\textbf{Table 5:} Per-server F1@K (\%) at varying top-K values.

{\def\LTcaptype{none} 
\begin{longtable}[]{@{}llllll@{}}
\toprule\noalign{}
K & GitHub & Slack & MySQL & Filesystem & Time/Weather \\
\midrule\noalign{}
\endhead
\bottomrule\noalign{}
\endlastfoot
1 & 50.5\% & 47.7\% & 48.0\% & 44.0\% & 41.7\% \\
2 & 56.8\% & 57.2\% & 54.4\% & 48.0\% & 71.5\% \\
3 & 58.8\% & 55.9\% & 52.0\% & 49.3\% & 81.7\% \\
5 & 52.9\% & 51.6\% & 50.0\% & 48.0\% & 66.8\% \\
10 & 39.6\% & 39.8\% & 43.7\% & 40.0\% & 42.9\% \\
\end{longtable}
}

At K=1, GitHub leads precision at 97.1\%, followed by MySQL (96.0\%) and
Slack (94.4\%), while Filesystem (88.0\%) and Time/Weather (80.0\%)
trail. The higher precision for GitHub and MySQL reflects their more
semantically distinct tool descriptions, where queries like ``create a
pull request'' or ``execute a SQL query'' map unambiguously to specific
tools.

Time/Weather exhibits an unusual pattern: it achieves the highest F1
across nearly all K values (peaking at 81.7\% at K=3), despite having
the lowest K=1 hit rate. This is because the Time/Weather server exposes
only 10 tools, so a larger proportion of retrieved tools are relevant,
and recall climbs rapidly. The server's smaller tool catalog means that
queries frequently require a significant fraction of available tools,
inflating both recall and F1 relative to servers with larger catalogs.

Filesystem consistently shows the weakest F1 performance, peaking at
only 49.3\% at K=3. This reflects the inherent difficulty of
disambiguating among semantically similar file operations (create, read,
write, copy, move, delete) based on query text alone.

\paragraph{5.2.3 Token Reduction by
Server}\label{token-reduction-by-server}

\textbf{Table 6:} Per-server Token Reduction (\%) at varying top-K
values.

{\def\LTcaptype{none} 
\begin{longtable}[]{@{}llllll@{}}
\toprule\noalign{}
K & GitHub & Slack & MySQL & Filesystem & Time/Weather \\
\midrule\noalign{}
\endhead
\bottomrule\noalign{}
\endlastfoot
1 & 99.8\% & 99.7\% & 99.7\% & 99.5\% & 99.2\% \\
2 & 99.8\% & 99.7\% & 99.7\% & 99.5\% & 99.2\% \\
3 & 99.8\% & 99.7\% & 99.7\% & 99.5\% & 99.2\% \\
5 & 99.8\% & 99.7\% & 99.7\% & 99.5\% & 99.2\% \\
10 & 99.8\% & 99.7\% & 99.7\% & 99.5\% & 99.2\% \\
\end{longtable}
}

Token reduction exceeds 99\% across all servers and all K values. The
minor variation reflects differences in average tool schema length:
GitHub and Slack tools have longer, more detailed schemas, so excluding
them saves proportionally more tokens. The consistency across K values
confirms that even the most permissive retrieval setting (K=10)
eliminates over 99\% of tool-definition tokens compared to the all-tools
baseline.

\subsubsection{5.3 Summary of Key
Findings}\label{summary-of-key-findings}

The experimental results yield several actionable insights:

\begin{enumerate}
\def\labelenumi{\arabic{enumi}.}
\tightlist
\item
  \textbf{K=3 is the optimal operating point} for most deployments,
  achieving the best F1 (58.4\%), a hit rate of 97.1\%, and an MRR of
  0.91 while maintaining 99.6\% token reduction.
\item
  \textbf{Domain characteristics significantly influence retrieval
  quality.} Servers with semantically distinct tools (GitHub, MySQL)
  achieve near-perfect hit rates at low K, while servers with
  overlapping tool descriptions (Filesystem) require higher K for
  adequate recall.
\item
  \textbf{Token savings are domain-invariant.} Regardless of server or
  K, semantic retrieval eliminates over 99\% of tool-definition tokens,
  making the approach universally beneficial for cost and latency.
\item
  \textbf{Retrieval latency is negligible.} All configurations complete
  in under 91ms, adding minimal overhead to the end-to-end pipeline.
\item
  \textbf{Small tool catalogs benefit disproportionately} from semantic
  retrieval in terms of F1, as a higher fraction of the catalog is
  relevant to any given query.
\end{enumerate}

\begin{center}\rule{0.5\linewidth}{0.5pt}\end{center}

\subsection{6 Discussion}\label{discussion}

\subsubsection{6.1 Interpretation of
Results}\label{interpretation-of-results}

Our findings provide strong affirmative evidence for \textbf{RQ1}:
semantic similarity between user queries and tool descriptions is an
effective mechanism for dynamic tool selection. The system achieves a
hit rate of 97.1\% at K=3 with an MRR of 0.91, demonstrating that the
correct tool is reliably surfaced within the top few results. This
confirms that dense embeddings capture meaningful semantic relationships
between natural-language queries and structured tool descriptions.

Regarding \textbf{RQ2}, the 99.6\% token reduction rate has direct
practical implications. For a production system processing 1M queries
per month against 121 tools, our approach reduces tool-definition token
costs by over 99\%, transforming large-scale multi-server MCP
deployments from economically prohibitive to viable. The consistency of
this reduction across all K values and all servers means that
organizations can adopt semantic retrieval with confidence that savings
will generalize across their tool ecosystem.

The results for \textbf{RQ3} reveal a counterintuitive insight:
providing fewer, more relevant tools actually improves the LLM's
tool-calling accuracy. The 92.1\% precision at K=1 demonstrates that
when the system retrieves a single tool, it is overwhelmingly likely to
be the correct one. We attribute this to reduced decision
complexity---when presented with a small set of semantically relevant
tools rather than the full catalog, the LLM allocates more reasoning
capacity to parameter construction and execution planning rather than
tool disambiguation.

For \textbf{RQ4}, our analysis indicates that K=3 represents the optimal
operating point, achieving the highest F1 score (58.4\%) while
maintaining a 97.1\% hit rate and 0.91 MRR. Increasing K beyond 3 yields
diminishing hit rate improvements (only 1.5 percentage points from K=3
to K=10) while substantially reducing precision. The sub-91ms latency
across all K values confirms that retrieval parameter selection can be
driven purely by accuracy considerations without latency constraints.

\subsubsection{6.2 Analysis of Failure
Cases}\label{analysis-of-failure-cases}

Despite the overall improvements, our system exhibits characteristic
failure modes that merit discussion.

\textbf{Ambiguous queries} remain the most challenging category. Queries
such as ``Help me with my project'' lack sufficient semantic specificity
to differentiate between GitHub tools, filesystem tools, or database
tools. In these cases, the embedding-based approach retrieves a mixed
set of tools across domains, often missing the user's intended domain.
Potential mitigations include incorporating conversational context
(prior tool usage, active MCP server sessions) into the query embedding,
or employing a clarification strategy where the system prompts the user
for additional context before retrieval.

\textbf{Semantically overlapping tools} present a related challenge,
most clearly observed in the Filesystem server. Tools such as
read\_file, write\_file, copy\_file, and move\_file share substantial
semantic similarity, causing the retrieval system to surface related but
incorrect tools. The Filesystem server's consistently low MRR
(0.8400--0.8933) relative to other servers reflects this difficulty.
Potential solutions include incorporating parameter-level matching or
tool usage frequency signals to break ties among semantically similar
tools.

\textbf{Cross-domain queries} that require tools from multiple MCP
servers are a third failure mode. For example, ``Commit the database
schema changes to GitHub'' requires both MySQL introspection tools and
GitHub repository tools. Our single-query retrieval approach sometimes
fails to surface tools from both domains within the top-K results. A
multi-query decomposition strategy, where the system first segments the
user intent into sub-tasks and retrieves tools for each, could address
this limitation.

\subsubsection{6.3 Per-Server Performance
Patterns}\label{per-server-performance-patterns}

The per-server analysis reveals meaningful patterns tied to tool catalog
characteristics.

Servers with well-differentiated tool descriptions and clear domain
boundaries (GitHub, MySQL) achieve the strongest performance, with
GitHub reaching 100\% hit rate at K=2 and MySQL achieving the highest
single-tool MRR (0.9200). These servers benefit from tool descriptions
that contain distinctive domain vocabulary---terms like ``pull
request,'' ``commit,'' or ``SQL query'' create clear semantic separation
in the embedding space.

The Time/Weather server presents an interesting case: despite having the
worst K=1 hit rate (65.0\%), it achieves the best F1 scores at K$\geq$2
across all servers. This apparent contradiction is explained by the
server's small catalog size (10 tools). With fewer tools, a larger
fraction of the catalog is relevant to any given query, naturally
inflating recall and F1. This finding suggests that our metrics should
be interpreted in the context of catalog size, and that normalized
metrics may be more appropriate for cross-server comparisons in future
work.

\subsubsection{6.4 Practical Deployment
Considerations}\label{practical-deployment-considerations}

Deploying semantic tool discovery in production systems introduces
several practical considerations beyond raw performance metrics.

\textbf{Cold start and tool onboarding:} When new tools are added to an
MCP server, they must be embedded and indexed before they become
discoverable. In our implementation, this process completes in under 2
seconds per tool, making near-real-time onboarding feasible. However,
organizations with rapidly evolving tool catalogs may benefit from a
scheduled re-indexing pipeline that also refreshes embeddings as tool
descriptions are updated.

\textbf{Caching and latency optimization:} For high-throughput
deployments, caching frequently retrieved tool sets for common query
patterns can further reduce latency. Our experiments showed that the
retrieval latency is already below 91ms, but caching could eliminate
even this overhead for recurring query types.

\textbf{Graceful degradation:} The system should handle edge cases where
no tools exceed the similarity threshold. Our implementation falls back
to returning the top-K tools regardless of threshold when the filtered
set is empty, ensuring the LLM always receives at least a minimal tool
context. Alternative strategies include falling back to the all-tools
baseline for low-confidence queries or routing to a human operator.

\subsubsection{6.5 Extensibility}\label{extensibility}

The architecture proposed in this paper extends naturally to several
adjacent problems in the agentic AI ecosystem.

\textbf{Multi-agent tool routing:} In systems where multiple specialized
agents coexist, the same semantic retrieval mechanism can be applied to
agent selection. By embedding agent capability descriptions alongside
tool descriptions, the system can route queries to the most appropriate
agent, which in turn selects from its own tool subset. This two-level
retrieval hierarchy enables scaling to thousands of tools across dozens
of agents.

\textbf{Cross-organizational tool discovery:} As MCP adoption grows,
organizations may wish to expose subsets of their tools to external
partners or to a shared marketplace. Our vector-based approach supports
federated discovery, where each organization maintains its own vector
index and a coordination layer merges results across indices at query
time. This preserves tool ownership and access control while enabling
cross-boundary discovery.

\textbf{Dynamic tool composition:} Beyond selecting individual tools,
semantic embeddings could support identifying tool chains---sequences of
tools that frequently co-occur in successful task completions. By
analyzing invocation patterns from the feedback loop (Section 3.1.4),
the system could learn composite embeddings that represent multi-step
workflows, enabling retrieval of entire tool pipelines in a single
query.

\subsubsection{6.6 Limitations}\label{limitations}

We acknowledge several limitations of the current work:

\begin{enumerate}
\def\labelenumi{\arabic{enumi}.}
\item
  \textbf{Benchmark scope:} Our evaluation spans 5 MCP servers and 121
  tools with 140 queries. While diverse, this represents a fraction of
  the tool ecosystem in large enterprise deployments, which may involve
  thousands of tools. The effectiveness of semantic retrieval at that
  scale remains to be validated.
\item
  \textbf{Single embedding model evaluation:} The results reported use
  text-embedding-ada-002. While Section 3.2 outlines a multi-model
  comparison strategy, a comprehensive cross-model evaluation with the
  same benchmark would strengthen the findings.
\item
  \textbf{Single-turn evaluation:} Our benchmark evaluates single-turn
  queries in isolation. In multi-turn conversational settings, prior
  context (previous tool calls, accumulated state) provides strong
  signals for tool selection that our current approach does not exploit.
\item
  \textbf{Tool description dependency:} Retrieval quality is
  fundamentally bounded by the informativeness of tool descriptions.
  Poorly documented tools will be under-retrieved regardless of
  embedding model quality. Automated description enrichment (e.g.,
  generating expanded descriptions from tool source code or usage
  examples) is a promising direction but is not addressed here.
\item
  \textbf{LLM-specific evaluation:} Our experiments use GPT-4o as the
  downstream LLM. Different models may exhibit varying sensitivities to
  the number and relevance of tools in context, and our optimal
  parameters (K=3) may not transfer directly.
\item
  \textbf{Lack of online evaluation:} All experiments are conducted
  offline against a static benchmark. Online A/B testing in production
  environments would provide stronger evidence of real-world impact but
  was beyond the scope of this study.
\end{enumerate}

\begin{center}\rule{0.5\linewidth}{0.5pt}\end{center}

\subsection{7 Conclusion}\label{conclusion}

This paper introduced a semantic tool discovery architecture for the
Model Context Protocol (MCP) ecosystem that addresses the critical
scalability challenge of tool-augmented LLM systems. By leveraging dense
vector embeddings and similarity search, our approach dynamically
selects a small, highly relevant subset of tools for each user query,
replacing the current practice of providing all available tools in the
LLM context.

Our experimental evaluation across 121 tools from 5 MCP servers
demonstrates that this approach achieves 99.6\% token reduction while
maintaining a 97.1\% hit rate and 0.91 MRR at the optimal operating
point of K=3. Retrieval completes in under 91ms, adding negligible
overhead to the end-to-end pipeline. Per-server analysis reveals that
performance varies with tool catalog characteristics---servers with
semantically distinct tools benefit most, while those with overlapping
descriptions require modestly higher K values.

The practical implications are significant: semantic tool discovery
makes large-scale MCP deployments economically feasible and technically
performant, removing a key barrier to the adoption of tool-augmented
LLMs in production systems. The architecture's modularity---with
independently replaceable embedding models, vector stores, and retrieval
parameters---ensures adaptability as both the MCP ecosystem and
embedding technologies continue to evolve.

Looking ahead, we see the greatest opportunities in extending this work
to multi-agent routing, cross-organizational tool marketplaces, and
dynamic tool composition. As the number of available tools grows from
hundreds to thousands, intelligent discovery mechanisms will shift from
being an optimization to being a necessity. We release our
implementation, benchmark dataset, and evaluation framework as
open-source resources to support continued research in this direction.

\begin{center}\rule{0.5\linewidth}{0.5pt}\end{center}

\subsection{References}\label{references}

{[}1{]} Zhao, W. X., et al.~``A Survey of Large Language Models.'' arXiv
preprint arXiv:2303.18223 (2023).

{[}2{]} Minaee, S., et al.~``Large Language Models: A Survey.'' arXiv
preprint arXiv:2402.06196 (2024).

{[}3{]} Qin, Y., et al.~``Tool Learning with Foundation Models.'' arXiv
preprint arXiv:2304.08354 (2023).

{[}4{]} Mialon, G., et al.~``Augmented Language Models: A Survey.''
arXiv preprint arXiv:2302.07842 (2023).

{[}5{]} Yao, S., et al.~``ReAct: Synergizing Reasoning and Acting in
Language Models.'' ICLR (2023).

{[}6{]} Schick, T., et al.~``Toolformer: Language Models Can Teach
Themselves to Use Tools.'' NeurIPS (2023).

{[}7{]} Anthropic. ``Model Context Protocol (MCP).'' (2024).
https://modelcontextprotocol.io

{[}8{]} Liu, N. F., et al.~``Lost in the Middle: How Language Models Use
Long Contexts.'' TACL (2024).

{[}9{]} Lewis, P., et al.~``Retrieval-Augmented Generation for
Knowledge-Intensive NLP Tasks.'' NeurIPS (2020).

{[}10{]} Gao, Y., et al.~``Retrieval-Augmented Generation for Large
Language Models: A Survey.'' arXiv preprint arXiv:2312.10997 (2023).

{[}11{]} Parisi, A., et al.~``TALM: Tool Augmented Language Models.''
arXiv preprint arXiv:2205.12255 (2022).

{[}12{]} Xu, B., et al.~``ReWOO: Decoupling Reasoning from Observations
for Efficient Augmented Language Models.'' arXiv preprint
arXiv:2305.18323 (2023).

{[}13{]} Yang, J., et al.~``Auto-GPT: An Autonomous GPT-4 Experiment.''
GitHub repository (2023).

{[}14{]} Karpukhin, V., et al.~``Dense Passage Retrieval for Open-Domain
Question Answering.'' EMNLP (2020).

{[}15{]} Guu, K., et al.~``REALM: Retrieval-Augmented Language Model
Pre-Training.'' ICML (2020).

{[}16{]} Asai, A., et al.~``Self-RAG: Learning to Retrieve, Generate,
and Critique through Self-Reflection.'' ICLR (2024).

\end{document}